\documentclass[twocolumn,aps,prb,amsmath,floatfix,showpacs]{revtex4}
\usepackage{graphicx}

\begin{document}
\title{Buried topological edge state associated with interface between topological 
band insulator and Mott insulator} 
\author{H. Ishida$^{1}$ and A. Liebsch$^2$}
\affiliation{$^1$College of Humanities and Sciences, Nihon University, Tokyo, 
              156-8550, Japan\\
$^2$Peter Gr\"unberg Institute and Institute of Advanced Simulations,
Forschungszentrum J\"ulich, 52425 J\"ulich, Germany}
\date{\today}

\begin{abstract}
The electronic structure at the interface between a topological band insulator 
and a Mott insulator is studied within layer dynamical mean field theory. To 
represent the bulk phases of these systems, we use the generalized Bernevig-Hughes-Zhang 
model that incorporates the Hubbard-like onsite Coulomb energy $U$ in addition to
the spin-orbit  coupling term that causes band inversion.
The topological and Mott insulating phases are 
realized by appropriately choosing smaller and larger values of $U$, respectively. 
As expected, the interface is found to be metallic because of the localized 
edge state. When the Coulomb energy in the Mott insulator is close to the critical value, 
however, this edge state exhibits its largest amplitude deep within the Mott insulator 
rather than at the interface. This finding corresponds to a new type of proximity effect 
induced by the neighboring topological band insulator and demonstrates that, as a result 
of spin-orbit coupling within the Mott insulator, several layers near the interface convert 
from the Mott insulating phase to a topological insulating phase. Moreover, we argue that 
the ordinary proximity effect, whereby a Kondo peak is induced in a Mott insulator by
neighboring metallic states, is accompanied by an additional reverse proximity effect, by
which the Kondo peak gives rise to an enhancement of the density of states
in the neighboring metallic layer.     
\end{abstract}

\maketitle

\section{Introduction}
\label{sec_1}

The role of Coulomb correlations in topological band insulators has recently received 
wide attention.\cite{Hohenadler} As a result of spin-orbit coupling, the band structure
of these materials exhibits charge excitation spectra whose physical characteristics 
can depend strongly on the Coulomb interaction. Iridium compounds, such as Na$_2$IrO$_3$
and A$_2$Ir$_2$O$_7$ (A=Pr,Eu) have recently been proposed as materials in which the 
interplay of spin-orbit interaction and electronic correlation effects might be important.
\cite{Shitade,Pesin} The phase diagram of a prototype bulk system of this kind was 
studied by Yoshida {\it et al.}\cite{Yoshida:12} within an extension of the two-band model 
proposed by Bernevig, Hughes, and Zhang (BHZ),\cite{BHZ} where an onsite Hubbard term is 
included to account for correlations. It was shown that, in the paramagnetic limit, such 
a system exhibits a first-order quantum phase transition, where the weakly correlated 
phase corresponds to a topological band insulator and the strongly correlated phase to a 
Mott insulator. 

At surfaces of topological insulators, metallic edge states may exist which are protected 
against perturbations associated with impurities and other interactions that do not break 
the time-reversal symmetry of the system.\cite{Hasan,Qi,Wu:2006,Xu,Yamaji} 
Because of these unique properties, heterostructures involving topological band insulators
\cite{Hutasoit,Yang,Wang,Ueda:13,Rauch,Essert} are presently of great interest since 
they might be relevant for future technological applications. For instance, as shown by Ueda 
{\it et al.},\cite{Ueda:13} the interface of a topological band insulator and a Mott insulator 
also exhibits an edge state which maintains its helical characteristics within the Mott insulator. 
Moreover, the quasi-particle properties and depth profile of this state within the Mott insulator 
depend strongly on the local Coulomb energy.         
       
\begin{figure}[t] 
\begin{center}
\includegraphics[width=7.5cm,height=5.5cm]{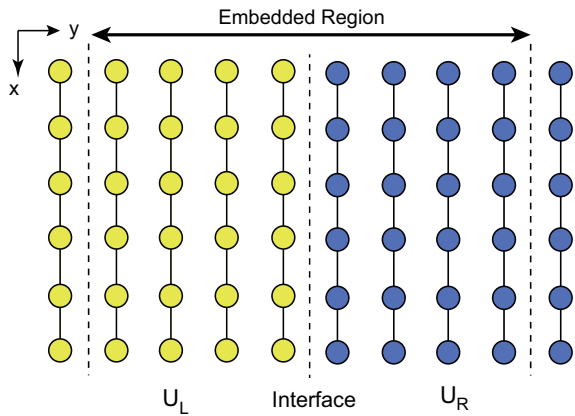}
\end{center}
\caption{\label{Fig1} (Color online)
One-dimensional interface between two-dimensional topological band and Mott insulators.
The Coulomb energies on the left and right side of the interface are defined as $U_L$
and $U_R$, respectively. The electronic properties in the embedding region are calculated
self-consistently within the layer DMFT. The properties of the asymptotic bulk regions
are taken into account via embedding potentials.}
\end{figure}

In the present work we study the role of electronic correlations at the interface between 
a topological insulator and a Mott insulator. The important difference between our approach 
and the one by Ueda {\it et al.}\cite{Ueda:13} is that we include spin-orbit coupling also 
within the Mott insulator. Thus, with decreasing Coulomb energy, the Mott insulator does 
not become a metal but a topological band insulator. The electronic properties in the vicinity 
of the interface are treated self-consistently by using the layer dynamical mean field theory 
(DMFT).\cite{DMFT,Potthoff,Okamoto,Helmes,Koga,Ishida:09,Gorelik}
Local many-body interactions are evaluated via finite-temperature exact diagonalization (ED).
\cite{Caffarel:94,Perroni:07,Liebsch:12}  Separate single-site DMFT calculations are performed 
for the asymptotic semi-infinite bulk regions. Their influence on the
interface region is taken into account via complex embedding potentials.\cite{Inglesfield:01}
For simplicity, a square lattice geometry is used, as illustrated in
Fig.\ \ref{Fig1}.\cite{Yoshida:12,Tada:12,Ueda:13} 
To incorporate spin-orbit interactions, the generalized two-orbital BHZ model is used, which
includes  the site-dependent Coulomb energy as well as the interorbital hybridization.

The main result of this study is the displacement of the edge state from the interface 
toward the interior of the Mott insulator when the local Coulomb energy on the corresponding  
side of the heterostructure is near the
critical value for the Mott transition. The edge state is then buried deeply within the 
Mott insulator so that the boundary between the band and Mott insulating phases no longer 
coincides with the physical interface of the two constituents of the heterostructure. 
The origin of this novel proximity effect is the fact that the Mott transition is first order. 
As a consequence, within the coexistence region topological band and Mott insulating
phases compete. Their relative stability depends sensitively on temperature and 
local Coulomb energy. Moreover, because of the 
penetration of the edge state wave function into the Mott insulator, near the interface 
the properties of the nominal Mott insulator are also influenced by the presence of the 
neighboring topological band insulator. As a result, the Mott insulating phase within
a certain depth can be converted into the more stable topological band insulating phase. 
The edge state is thereby displaced away from the physical interface which becomes 
the boundary between two weakly and strongly correlated topological band insulators. 
As the Coulomb energy in the Mott insulator increases beyond the coexistence region, 
the edge state is localized again at the interface. Its depth within the Mott insulator 
diminishes as the Mott gap increases. Below the coexistence domain, the edge disappears 
since the heterostrucure then consists of two correlated topological band insulators. 

We emphasize that the displacement of the edge state away from the interface of the 
heterostructure is a consequence of the interorbital hybridization (spin-orbit coupling)
in the Mott insulator. In the absence of spin-orbit coupling, the coexistence domain 
involves trivial metallic and insulating phases so that a topological band insulating 
solution does not occur. Thus, if an edge state exists, it is localized at the physical 
interface.

We also demonstrate that the ordinary proximity effect,\cite{Helmes,Ishida:09,Zenia,Borghi}  
i.e., the appearance of a Kondo peak in a Mott insulator due to neighboring metallic states, 
is accompanied by an secondary reverse proximity effect, as a result of which the Kondo peak 
leads to an increase of the density of states (DOS) in the neighboring metallic layer.

The outline of this paper is as follows. Section II presents the main aspects of the 
theoretical approach. In particular, we introduce the generalized BHZ two-band model which 
provides the basis for the topological band as well as Mott insulating phases. Also, the 
embedding scheme is described in which the effect of the asymptotic bulk materials on 
either side of the interface are taken into account via complex local potentials. 
Finally, the inhomogeneous layer DMFT is outlined, as well as the finite-temperature 
exact diagonalization scheme for the treatment of local many-body interactions.
Section III provides the discussion of the results. We first present the phase diagram 
of the asymptotic bulk materials and illustrate the edge state at the solid-vacuum interface.
The main part discusses the electronic properties of the interface between topological 
band and Mott insulators, in particular, the location of the edge state as a function 
of the Coulomb energy within the Mott insulator.         

\section{Theory}
\label{sec_2}

\subsection{Method}
\label{sec_2A}

We consider a one-dimensional interface between a two-dimensional (2-D) topological
band insulator (BI) and a 2-D Mott insulator (MI), which occupy the left and right
half-space, respectively. The $x$ direction is parallel to the interface, while the $y$
axis, which points from left to right, is chosen as the interface normal.
To represent the semi-infinite systems on both sides, we employ the generalized
Bernevig-Hughes-Zhang model,
\begin{eqnarray}
\hat{H}&=& \hat{H}_{\rm bhz} + \hat{H}_{\rm int} \nonumber\\
       &=& (\hat{H}_0+\hat{H}_{\rm so}) + \hat{H}_{\rm int}. \label{eq1}
\end{eqnarray}
The first term in the second line of Eq.\ (\ref{eq1}),
\begin{equation}
\hat{H}_0=\sum_{p,\alpha,\sigma}\left(\epsilon_\alpha-\frac{U_y}{2}\right) \hat{n}_{p \alpha \sigma}
+\sum_{\langle p,q\rangle,\alpha,\sigma} t_\alpha c^\dagger_{p \alpha \sigma}
c_{q \alpha \sigma}, \label{eq2}
\end{equation}
represents two tight-binding bands originating from two orbitals,
where $c^\dagger_{p \alpha \sigma}$  ($c_{p \alpha \sigma}$) creates (annihilates)
an electron with orbital $\alpha=1, 2$ in spin state $\sigma=1\ (\uparrow), -1\ (\downarrow)$ on
a 2-D square lattice point at $p=(x,y)$, with $x$ and $y$ giving its $x$ and $y$ positions,
respectively. In Eq.\ (\ref{eq2}), $\epsilon_\alpha$ and $t_\alpha$ are the site
energy and nearest-neighbor hopping integral for orbital $\alpha$,
$U_y$ is the Coulomb energy, which will be described below, 
$\hat{n}_{p \alpha \sigma}=c^\dagger_{p \alpha \sigma} c_{p \alpha \sigma}$ denotes the orbital
occupation, and the summation over $p$ and $q$ in the second term is taken over nearest-neighbor
lattice-point pairs. 
The second term in the second line of Eq.\ (\ref{eq1}), which arises from
spin-orbit coupling and is responsible for the opening of a topological energy band gap, reads
\begin{equation}
\hat{H}_{\rm so}=t_{12} \sum_{\langle p,q\rangle,\sigma} i\sigma\ \left[e^{i\theta \sigma}
c^\dagger_{p 2 \sigma} c_{q 1 \sigma}
+e^{-i\theta \sigma}c^\dagger_{p 1 \sigma} c_{q 2 \sigma}\right],\label{eq3}
\end{equation}
where $\theta$ specifies the hopping direction measured relative to the $x$ axis
($\theta=0$ and $\pi/2$ correspond to the hopping to the positive $x$ and $y$ directions,
respectively). The last term in Eq.\ (\ref{eq1}),
\begin{equation}
\hat{H}_{\rm int} = \sum_{p,\alpha} U_y\ \hat{n}_{p \alpha \uparrow} \hat{n}_{p \alpha \downarrow},
\label{eq4}
\end{equation}
expresses the onsite Coulomb repulsion between electrons with opposite spin in the same orbital
$\alpha$. We assume that the Coulomb energy can vary with lattice layers, while it is
constant within the same layer.
It should be noted that the term $-\frac{1}{2}U_y\hat{n}_{p \alpha \sigma}$ in Eq.\ (\ref{eq2})
ensures that the system becomes electron-hole symmetric when chemical potential $\mu$
is chosen as $\mu=0$. 

Yoshida {\it et al}.\cite{Yoshida:12} studied the effect of strong Coulomb correlations on
a topological band insulator by applying single-site DMFT\cite{DMFT}
to periodic 2-D bulk systems described by the same generalized BHZ model.
It was shown that the system undergoes a quantum phase transition from a topological band insulator 
to a Mott insulator when one increases the onsite Coulomb energy $U$, while keeping the system nonmagnetic.
As schematically illustrated in Fig.\ \ref{Fig2},
the phase transition is of first order and exhibits hysteresis behavior, i.e., both topological 
band and Mott insulating solutions are found if $U$ is within the coexistence region $[U_{c1},U_{c2}]$. The width
of this region decreases with increasing temperature $T$ and vanishes at a critical value $T_c$.

In the present work, both the topological band insulator on the left half-space and the Mott 
insulator on the right half-space are represented by the generalized BHZ model as described above.
As indicated in Fig.\ \ref{Fig1}, the layer dependent Coulomb energy, $U_y$, is set
to be $U_L$ and $U_R$ in the left and right half-spaces, respectively, where
$U_L < U_{c2}$ and $U_R>U_{c1}$ (see Fig.~2 below).
We note that our model differs from that in the recent work of Ueda
{\it et al}.\cite{Ueda:13}, in which the Mott insulator was represented by two independent
Hubbard bands without the spin-orbit coupling term defined in Eq.\ (\ref{eq3}).

\subsection{Embedding potential}
\label{sec_2B}

We calculate the finite-temperature Green's function of the interface between two
semi-infinite systems by using the layer DMFT technique.\cite{Potthoff} A finite number 
of lattice layers in the interface region is treated explicitly, whereas the 
effect of the outer regions is taken into account via the embedding potentials\cite{Inglesfield:01}
which include correlation effects in the bulk region.\cite{Ishida:09}
As we consider nonmagnetic solutions in the present work, the Green's function and other
quantities are diagonal with respect to spin. In the following, we show only the up-spin
component of the equations and omit spin indices for simplicity.   

By introducing the wave number in the $x$ direction, $k_x$, the Green's function in the
embedded region is given as
\begin{eqnarray}
G_{p\alpha,p'\alpha'}(i\omega_n) =  \int_{-\pi}^{\pi} \frac{dk_x}{2\pi}
 e^{i k_x (x-x')} \hspace{2cm} \nonumber \\
\times \langle y\alpha\mid [i\omega_n+\mu-\hat{H}_{\rm emb}(k_x,i\omega_n)]^{-1}
\mid y'\alpha'\rangle,\label{eq5}
\end{eqnarray}
where the embedding Hamiltonian in the mixed representation, $\hat{H}_{\rm emb}$, is a
$2N\times 2N$ matrix with $N$ being the number of the embedded lattice layers. 
It consists of four terms:
\begin{eqnarray}
\hat{H}_{\rm emb}(k_x,i\omega_n) &=& \hat{H}^N_{\rm bhz}(k_x)+\hat{\Sigma}(i\omega_n)\nonumber\\
&+&\hat{s}^L(k_x,i\omega_n)+\hat{s}^R(k_x,i\omega_n),\label{eq6}
\end{eqnarray}
where $\hat{H}^N_{\rm bhz}(k_x)$ denotes the one-electron part of the Hamiltonian in Eq.\ (\ref{eq1}).
The superscript $N$ emphasizes the fact that $\hat{H}^N_{\rm bhz}$ is a $2N\times2N$ matrix
for an isolated slab. The second term in Eq.\ (\ref{eq6}) is the Coulomb self-energy. Within single-site
DMFT, it is layer diagonal and $k_x$-independent:
\begin{equation}
\langle y\alpha \mid \hat{\Sigma} \mid y'\alpha'\rangle
=\Sigma_{\alpha \alpha'}(y,i\omega_n) \delta_{y,y'}. \label{eq7}
\end{equation}
The last two terms in Eq.\ (\ref{eq6}) are the embedding potentials. Since $\hat{H}_{\rm bhz}$
in Eq.\ (\ref{eq1}) includes only nearest-neighbor hopping terms, $\hat{s}^L$ ($\hat{s}^R$) is
non-vanishing only when both layer indices are equal to $y_L$ ($y_R$), the outermost layer of
the embedded slab region on the left (right) hand side. Thus, they are written as
\begin{eqnarray}
\langle y\alpha \mid \hat{s}^L \mid y'\alpha'\rangle
&=&s^L_{\alpha \alpha'}(k_x,i\omega_n) \delta_{y,y_L} \delta_{y',y_L}, \label{eq8}\\
\langle y\alpha \mid \hat{s}^R \mid y'\alpha'\rangle
&=&s^R_{\alpha \alpha'}(k_x,i\omega_n) \delta_{y,y_R} \delta_{y',y_R}. \label{eq9}
\end{eqnarray}

We now explain how we can derive the embedding potential for the left-hand side.
We assume that the electronic structure in the half-space to the left of the embedded slab
region converges to that of the bulk crystal with Coulomb repulsion energy $U_L$. Thus, within
the single-site approximation, the Coulomb self-energy of all layers is assumed to be
identical to that in the interior of the bulk with $U_L$, $\hat{\Sigma}_L(i\omega_n)$. This quantity 
is a $2\times2$ matrix in orbital space. 
Now, let us consider the Green's function of the semi-infinite solid in
which the self-energies of all layers are equal to $\hat{\Sigma}_L(i\omega_n)$. We extract
from this Green's function a $2\times2$ matrix spanned by the two orbital components
of the outermost surface layer, which is denoted by $g^L_{\alpha \alpha'}(k_x,i\omega_n)$.
The embedding potential, i.e., the $2\times 2$ matrix appearing on the right hand side of
Eq.\ (\ref{eq8}), is then given by\cite{Ishida:09}
\begin{equation}
\hat{s}^L(k_x,i\omega_n) = \hat{t}_{+}\ \hat{g}^L(k_x,i\omega_n)\ \hat{t}_{-},\label{eq10}
\end{equation}
with
\begin{equation}
\hat{t}_{+}=
\begin{pmatrix}
t_1 & t_{12}\\
-t_{12} & t_2
\end{pmatrix},\
\hat{t}_{-}=
\begin{pmatrix}
t_1 & -t_{12}\\
t_{12} & t_2
\end{pmatrix},\label{eq11}
\end{equation}
where $t_{+}$ ($t_{-}$) is the transfer matrix for electrons which hop between two
nearest-neighbor lattice layers toward the positive (negative) $y$ direction.

In order to obtain $\hat{g}^L$ in Eq.\ (\ref{eq10}), we use the following trick.
We add one additional lattice layer having the bulk self-energy $\hat{\Sigma}_L$ on
top of the semi-infinite substrate expressed by the embedding potential $\hat{s}^L$.
Then, the $2\times2$ surface Green's function of the resultant new semi-infinite solid
is given by
\begin{equation}
\hat{g}^L=\left[i\omega_n+\mu-\hat{H}^{N=1}_{\rm bhz}(k_x)-\hat{\Sigma}_L(i\omega_n)
-\hat{s}^L\right]^{-1},\label{eq12}
\end{equation}
where $\hat{H}^{N=1}_{\rm bhz}(k_x)$ is given by
\begin{equation}
\hat{H}^{N=1}_{\rm bhz}(k_x)=
\begin{pmatrix}
\epsilon_1-\frac{U_L}{2}+ 2 t_1 \cos k_x & 2 t_{12} \sin k_x \\
2 t_{12} \sin k_x & \epsilon_2-\frac{U_L}{2} + 2 t_2 \cos k_x
\end{pmatrix}.\label{eq13}
\end{equation}
We now make use of the fact that the semi-infinite solid obtained by putting one
bulk lattice on top of the semi-infinite bulk system is exactly the same as that 
without the additional lattice layer. Thus, $\hat{g}^L$ on the right-hand side of Eq.\ (\ref{eq10})
must coincide with $\hat{g}^L$ calculated by Eq.\ (\ref{eq12}). Thus, by inserting $\hat{g}^L$
in Eq.\ (\ref{eq12}) into the right-hand side of Eq.\ (\ref{eq10}), one obtains a set of
equations to determine the three independent elements of the embedding potential,
$s^L_{\alpha \alpha'}$ for given $k_x$ and $\omega_n$. In contrast to one-band models,
for which one can derive an analytical expression of the embedding potential from a
quadratic equation obtained by following the same procedure as described above,\cite{Ishida:09}
the embedding potential for the present two-band model can be computed only numerically.
To determine $\hat{s}^L$, one needs the bulk self-energy $\Sigma_L(i\omega_n)$ in
Eq.\ (\ref{eq12}), which can be determined from an independent DMFT calculation for
the 2-D bulk system with Coulomb repulsion energy $U_L$, before evaluating the interface
properties.

Similarly, the embedding potential for the right-hand side can be derived by combining
the two following equations.
\begin{equation}
\hat{s}^R(k_x,i\omega_n) = \hat{t}_{-}\ \hat{g}^R(k_x,i\omega_n)\ \hat{t}_{+},\label{eq14}
\end{equation}
\begin{equation}
\hat{g}^R=\left[i\omega_n+\mu-\hat{H}^{N=1}_{\rm bhz}(k_x)-\hat{\Sigma}_R(i\omega_n)
-\hat{s}^R\right]^{-1},\label{eq15}
\end{equation}
where $\Sigma_R(i\omega_n)$ denotes the Coulomb self-energy of the 2-D bulk system
with Coulomb interaction $U_R$.

\subsection{DMFT equation and exact diagonalization}
\label{sec_2C}

Starting from some initial self-energy matrix, one calculates the local components of the
lattice Green's function for each layer,
$g_{\alpha\alpha'}(y, i\omega_n)=G_{p\alpha,p\alpha'}(i\omega_n)$,
by using Eq.\ (\ref{eq5}). Then, the bath Green's function determining the Weiss mean-field
of layer $y$ is obtained by removing the local self-energy:
\begin{equation}
\hat{g}^0(y,i\omega_n) =\left[\hat{g}^{-1}(y,i\omega_n)+\hat{\Sigma}(y,i\omega_n)\right]^{-1},
\label{eq16}
\end{equation}
where $\hat{\Sigma}(y,i\omega_n)$ is the $2\times2$ matrix defined by Eq.\ (\ref{eq7}).

In the present work,
the quantum impurity problem is solved by making use of the exact diagonalization (ED)
formalism,\cite{Caffarel:94,Perroni:07,Liebsch:12} in which $\hat{g}^0(y,i\omega_n)$ is
approximated by a noninteracting Green's function of a finite cluster consisting of two
impurity levels with energy $E_\alpha$ coupled to $n_b$ bath orbitals with energy
$\epsilon_k$. Thus,
\begin{equation}
\hat{g}^0(y,i\omega_n) \approx \hat{g}^{cl,0}(y,i\omega_n)
=\left[i\omega_n+\mu-\hat{h}^{cl}(i\omega_n)\right]^{-1},\label{eq17}
\end{equation} 
with
\begin{equation}
\hat{h}^{cl}_{\alpha\alpha'}(i\omega_n)=E_\alpha\ \delta_{\alpha\alpha'}
+\sum_{k=1}^{n_b}\frac{v_{\alpha k} v_{k\alpha'}}{i\omega_n-\epsilon_k},\label{eq18}
\end{equation}
where $E_\alpha$, $\epsilon_k$ and $v_{\alpha k}$ are real fitting parameters chosen
such that the weighted sum of  $|\hat{g}^0-\hat{g}^{cl,0}|^2$ over a sufficiently large
Matsubara frequency range is minimized.\cite{Liebsch:12}
Then, after adding the onsite Coulomb repulsion terms Eq.\ (\ref{eq4}) to this $(2+n_b)$-level
cluster, the interacting Green's function of the cluster,  $\hat{g}^{cl}(y,i\omega_n)$, is
derived by combining the Arnoldi algorithm for computing the lowest eigenstates with the
Lanczos procedure for calculating the Green's function.\cite{Perroni:07}
Finally, the cluster self-energy is obtained from the equation:
\begin{equation}
\hat{\Sigma}^{cl}(y,i\omega_n)=\left[\hat{g}^{cl,0} (y,i\omega_n)\right]^{-1}   
-\left[\hat{g}^{cl} (y,i\omega_n)\right]^{-1}.\label{eq19}
\end{equation}

In the ED formalism, the cluster self-energy is assumed to be a physically reasonable
representation of the lattice self-energy. Thus,  $\hat{\Sigma}^{cl}(y,i\omega_n)$ is
used as the input self-energy   $\hat{\Sigma}(y,i\omega_n)$ in Eq.\ (\ref{eq5}) for the
next DMFT iteration. This procedure is iterated until the difference between the input
and output self-energy matrices for all the lattice layers in the embedded region becomes
sufficiently small. In the calculation presented in the next section, we use $n_b=8$
bath orbitals (4 per orbital), so that the total number of energy levels per cluster equals 10.  

\section{Results and discussion}
\label{sec_3}

In the present work, the parameters of the non-interacting part of the Hamiltonian
are chosen as $\epsilon_1=-1$, $t_1=-1$, $\epsilon_2=1$, $t_2=1$, and $t_{12}=0.5$.
The same parameter set was used previously in Ref.~\onlinecite{Yoshida:12}.
We consider only the electron-hole symmetric case with chemical potential $\mu=0$.
In the absence of correlations, the bulk bands extend from $-3$ to $+3$ and the band 
gap from $-1$ to $+1$.
The DMFT calculations are performed at a relatively small temperature: $T=1/\beta=0.01$.

\subsection{Bulk phase diagram}
\label{sec_3A}

\begin{figure}[t] 
\begin{center}
\includegraphics[width=7.5cm,height=5.0cm]{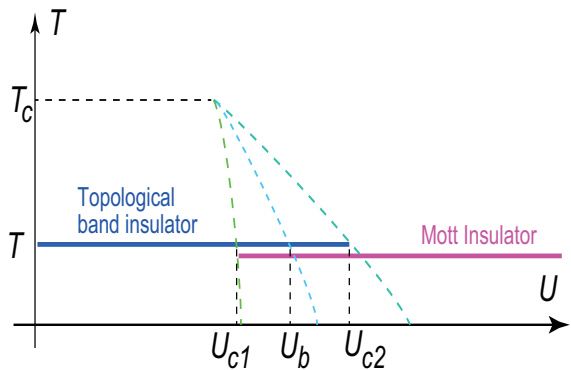}
\end{center}
\caption{\label{Fig2} (Color online)
Schematic bulk phase diagram of correlated topological band insulator, derived within generalized
BHZ two-orbital model and single-site DMFT. In the region limited by the lines $U_{c1}(T)$ and 
$U_{c2}(T)$, band and Mott insulating states may coexist. The true phase boundary defining the 
relative stability of these phases is indicated by the line $U_b(T)$. 
}\end{figure}

Figure \ref{Fig2} illustrates schematically the bulk phase diagram of the constituents of the present
heterostructure. At small $U$, one has a weakly correlated topological band insulator which 
corresponds to the system on the left-hand side in Fig.\ \ref{Fig1}. At large $U$, the system becomes
a Mott insulator which is taken to be the dominant phase on the right-hand side of the interface.      
We note that in the present model the Mott insulating phase is nontopological with a vanishing Chern
number, in contrast to the topological Kondo insulator.\cite{Dzero:10,Werner:12}  
The band gap in the band insulator also varies with $U$. It is largest in the noninteracting 
limit and gradually decreases with increasing $U$ until $U$ approaches 
$U_{c2}(T)$. This behavior corresponds to the usual band-narrowing effect, which has been 
discussed in previous work.\cite{Sentef,Grandi} As a result of local Coulomb interactions,
spectral weight within the region of the bulk bands is transferred to low energies and Hubbard 
bands appear at high energies. For the parameters specified above, we find that the coexistence 
region is limited by the boundaries $U_{c1}\approx 11.4$ and $U_{c2}\approx 13.3$. 
Because of the first-order nature of the Mott transition, various quantities, such as the orbital
polarization, the double occupancy of the subbands, the spectral weight at the chemical potential, 
etc. exhibit the usual hysteresis behavior (not shown here). An important aspect of the interface 
properties discussed in subsection \ref{sec_3D} is the fact that they may be used to determine the
relative stability of the band and Mott insulating phases in the coexistence region. For instance, at
$T=0.01$ we estimate $U_b\approx 13.0$.     

As will be shown below, an edge state appears at the interface of a weakly correlated topological
band insulator and a strongly correlated Mott insulator. The intriguing question then arises what 
happens to the edge state when the Coulomb energy in the Mott insulator lies within the coexistence 
region. Depending on the precise values of $U$ and $T$, the Mott insulating phase can become unstable 
and may therefore be converted into a band insulating phase. Before we discuss this case, we consider 
in the following two subsections (i) the behavior of the edge state at the bare surface of a topological 
band insulator and (ii) the properties of the interface between two topological insulators.          
These results serve a useful reference for the subsequent analysis of the interface with a Mott insulator

\subsection{Edge State at Solid-Vacuum Interface} 
\label{sec_3B}

\begin{figure}[t] 
\begin{center}
\includegraphics[width=7.5cm,height=5.0cm]{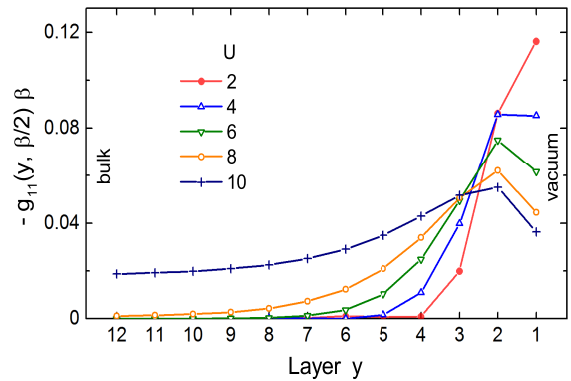}
\end{center}
\caption{\label{Fig3} (Color online)
Variation of edge state at bare surface of correlated topological band insulator as a function
layer index  $y$ for various Coulomb energies $U$, as calculated within inhomogeneous DMFT.
Plotted is the function $-\beta g_{11}(y,\beta/2)$,
defined in Eq.~(\ref{G}), which represents the partially integrated density of states within a
few $T$ of the chemical potential ($T=0.01$).
The embedding region at the surface comprises 12
atomic layers beyond which bulk behavior is assumed.
}\end{figure}

Figure \ref{Fig3} shows the spectral weight near the chemical potential at the surface of a correlated
topological band insulator as a function of distance from the surface for several values of $U$.
Plotted is the function $-\beta g_{11}(y,\beta/2)$ which provides a measure of the DOS
within a few $T$ of the chemical potential. This quantity is defined as
\begin{eqnarray}\label{G}
-\beta g_{\alpha\alpha}(y,\beta/2) &=& - \sum_n g_{\alpha\alpha}(y,i\omega_n)\, e^{-i\omega_n\beta/2}
      \nonumber\\
    &=& \pi \int^\infty_{-\infty} d\omega\, F(\omega) N_{\alpha}(\omega),
\end{eqnarray}
where $N_{\alpha}(\omega) = - \frac{1}{\pi} {\rm Im} g_{\alpha\alpha}(y,\omega+i\delta)$ is the
interacting DOS of subband $\alpha$ and the weight function $F$ is defined as:
$F(\omega) = 1/[2\pi T\cosh(\omega/(2T))]$. 
(The width of $F$ is about $5.3T$; its integrated weight is unity.)
As a result of particle-hole symmetry, $g_{11}(y,\beta/2)= g_{22}(y,\beta/2)$.

At the surface of a topological insulator, a metallic edge state connecting the bulk valence
and conduction bands appears. As the chemical potential is located at the middle of the
energy gap when the system is electron-hole symmetric, the edge state contributes to an
increase in DOS at $\mu$ for several surface layers, which is clearly seen in Fig.\ \ref{Fig3}. 
With increasing values of $U$, the gap in the topological band insulator diminishes due to 
correlation effects, so that the penetration depth of the edge state increases. The peak of 
the edge state at larger $U$ also shifts to the second layer (see below). At the same time,
the high-energy tails of $F(\omega)$ on both sides of $\omega=0$ start overlapping with the 
bulk bands due to band-gap narrowing. For $U>8$, this results in a rapid increase in the 
calculated values of $-\beta g_{11}(y,\beta/2)$ in the interior of the solid.

\begin{figure}[t] 
\begin{center}
\includegraphics[width=0.4\textwidth]{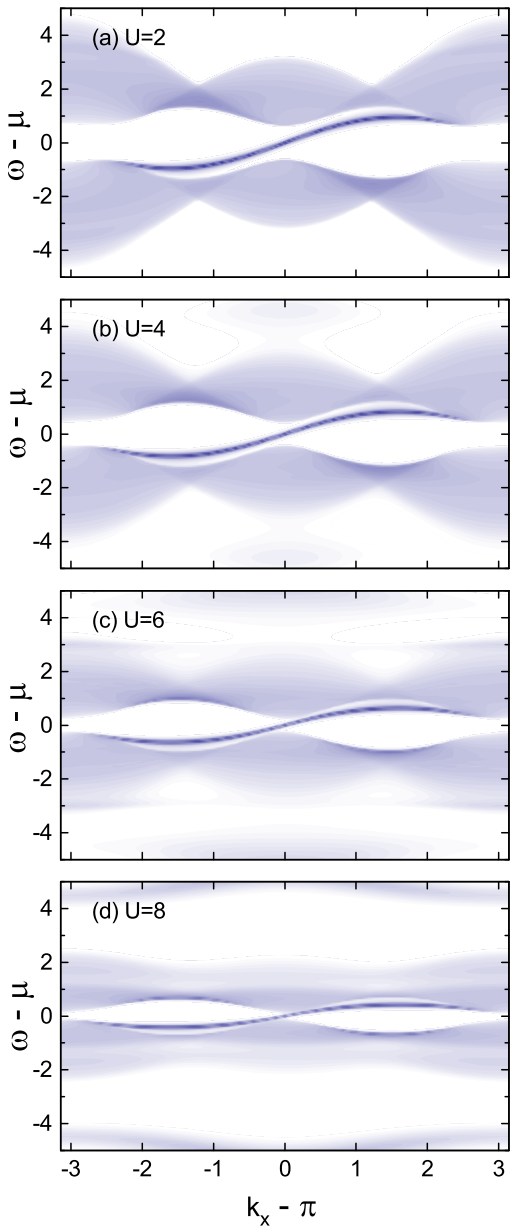}
\end{center}
\caption{\label{Fig4} (Color online)
Intensity plot of interacting DOS of orbital $\alpha=1$, $\sigma=1$ in surface layer of
semi-infinite topological insulator as a function of parallel momentum $k_x$ for several
values of $U$.
}\end{figure}

\begin{figure}[t] 
\begin{center}
\includegraphics[width=0.35\textwidth]{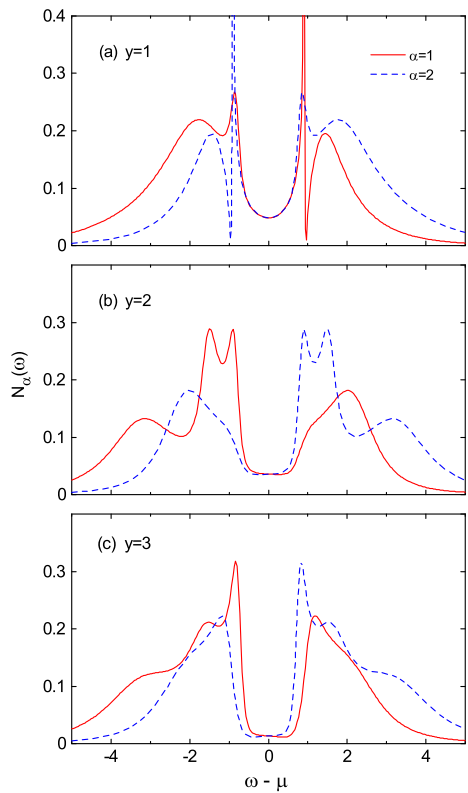}
\end{center}
\caption{\label{Fig5} (Color online)
Interacting DOS,  $N_{\alpha}(\omega)$,  in first three surface layers of topological insulator
at $U=2$. Solid (red) curves: orbital $\alpha=1$, dashed (blue) curves: $\alpha=2$. These spectra are
derived by extrapolating the lattice Green's function from Matsubara frequencies to
$\omega+i\gamma$ with a small imaginary energy $\gamma=0.05$ via the routine {\it ratint}.
The spectral weight near $\mu=0$ in panel (a) is due to the metallic edge state, while the DOS in
panel (c) approaches the one characteristic of the bulk band gap.
}\end{figure}

To illustrate the energy dispersion of the edge state with $k_x$ in the case of a free
surface, we show in Fig.\ \ref{Fig4} the $k_x$-resolved DOS of the first layer ($y=1$) for
various Coulomb energies. These results were derived by extrapolating the self-energy
from the Matsubara axis to real energies via the  routine {\it ratint}
(for details see Ref.~\onlinecite{Liebsch:12}) and then evaluating the integrand on the
right hand side of Eq.\ (\ref{eq5}).
One can clearly see a metallic edge state crossing the chemical potential $\mu=0$ at
$k_x=\pi$ with a positive group velocity. This indicates that the increase in DOS
seen in Fig.\ \ref{Fig3} for surface layers arises from the edge state around
$k_x=\pi$. The energy dispersion curve of the edge state for the down-spin (not shown)
is obtained by reflecting the one for the up-spin shown in Fig.\ \ref{Fig4} with respect to
$k_x=\pi$, so that the two dispersion curves with the opposite spins cross at $k_x=\pi$.

With increasing $U$, the dispersion of the edge state becomes flatter and the bulk band 
gap is reduced. Also, the width of the bulk bands decreases and Hubbard bands appear below 
and above the band region.
Since the weight function $F(\omega)$ in Eq.\ (\ref{G}) acts nearly like a $\delta$-function
at low $T$, the correlation induced band narrowing of the edge state gives rise to an increase
of $-\beta g_{11}(y,\beta/2)$. As shown in Fig.\ \ref{Fig3},
this is indeed the case for layers $y>2$. Remarkably, this increase of density for $y>2$ is
compensated by a corresponding decrease for layers $y=1$ and $2$, so that the layer-integrated 
weight of the edge state close to $\mu$ remains nearly constant. 

According to the dispersions shown in Fig.\ \ref{Fig4}, the DOS of the edge state in the surface layer 
has the typical shape of a one-dimensional tight-binding system, with a minimum at the center and 
logarithmic van Hove singularities at the band limits. This is illustrated in Fig.\ \ref{Fig5}(a) for $U=2$. 
Panels (b) and (c) indicate how this one-dimensional metallic spectral distribution converts
to the one of the topological band insulator as one moves away from the surface.             
We also note that the DOS at $\omega=\mu$ in Fig.\ \ref{Fig5} decays with increasing
$y$ more slowly than $-\beta g_{11}(y,\beta/2)$ in Fig.\ \ref{Fig3} owing to a small
imaginary energy $\gamma=0.05$ introduced in extrapolating the lattice Green's function.  
The peaks at larger energies are sensitive to details of the extrapolation procedure.
  
As will be seen in Subsection \ref{sec_3D}, the low-energy spectral distribution of the edge state 
of the bare surface shown in Fig. 5 differs qualitatively from the one of edge state at the 
interface between a topological band insulator and a Mott insulator. The main reason is the  
appearance of a Kondo peak in the Mott insulator and a new single-particle feature at $\mu$ 
at the surface of the topological insulator.        

\subsection{Interface between Topological Band Insulators}
\label{sec_3C}

\begin{figure}[t] 
\begin{center}
\includegraphics[width=7.5cm,height=10.0cm]{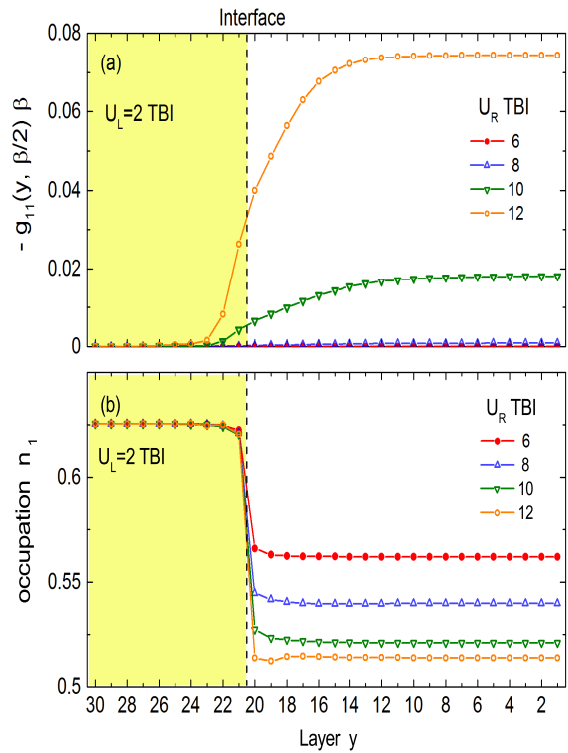}
\end{center}
\caption{\label{Fig6} (Color online)
(a) Partially integrated spectral weight, $-\beta g_{11}(y,\beta/2)$, and (b) occupancy $n_1$ of
orbital $\alpha=1$ as functions of layer index near the interface between two correlated topological
band insulators. The systems on the left and right sides have local Coulomb energies $U_L=2$ and
$U_R=6 \ldots 12$, respectively. The location of the interface is indicated by the dashed line.
The embedding region consists of 10 layers on the left and 20 layers on the right side of the interface.
The  asymptotic behavior is derived from DMFT calculations for the respective bulk materials.
}\end{figure}

Figure \ref{Fig6}(a) shows the partially integrated DOS $-\beta g_{11}(y,\beta/2)$ at the interface
between a weakly correlated topological band insulator ($U_L=2$) and more strongly correlated ones 
($U_R=6 \ldots 12$) as a function of layer index $y$. These curves vary monotonously across the 
interface between two asymptotic values of $-\beta g_{11}(y,\beta/2)$ which are the same as those 
in the interior of the semi-infinite solid shown in Fig.\ \ref{Fig3}.
As mentioned above, the asymptotic values starts growing  due to band narrowing
only for $U>8$. Therefore, the curves for $U=6$ and 8 in Fig.\ \ref{Fig6}(a) are practically constant
throughout the system, while the curves for $U=10$ and $12$ exhibit a smooth variation between the two
asymptotic values.  Evidently, there is no sign of a topological edge state at the interface. 

Figure \ref{Fig6}(b) shows the occupancy $n_1$ as a function of layer index $y$ for the three
systems shown in panel (a). As we consider electron-hole symmetric systems, the occupancies of the 
two orbitals are related by $n_2=1-n_1$. It is seen that the orbital polarization, i.e., $n_1-n_2$,
is strongly reduced in the right-hand system with increasing values of  Coulomb energy $U$.
(At the Mott transition, both orbitals become half-filled, i.e., $n_1 = n_2 = 0.5$.) Whereas the
variation of  $n_1$ occurs very rapidly within one or two layers near the interface, the 
variation of $-\beta g_{11}(y, \beta/2)$ for $U_R > 8$ is more gradual and comparable to the one
at the bare surface shown in Fig.\ \ref{Fig3}. This difference is related to the fact that, because 
of the finite width of the weight function $F(\omega)$, $g_{11}(y,\beta/2)$ is mainly sensitive
to the density of states of the low-energy bulk bands, in particular, at larger $U$, when the 
band gap shrinks.

\subsection{Interface between Topological Band and Mott Insulators}
\label{sec_3D}

We now discuss in more detail the properties of the edge state between a topological band insulator
($U_L=2$) and a Mott insulator whose Coulomb energy is close to the coexistence region.
To obtain the self-consistent solution of the layer-coupled DMFT equation, we adopt the
following procedure: On the left (right) boundary of the embedded region, we apply the embedding
potential for a semi-infinite solid in the topological insulator phase with $U_L=2$ (Mott insulator
phase with varying value of $U_R$), so that the physical states in the asymptotic regions are fixed.
Within the embedded region, the initial values of the cluster parameters in Eq.\ (\ref{eq18}),
which determine the initial Weiss mean-field for each layer, are taken to be the parameters representing
the bulk topological insulator with $U_L$ (Mott insulator with $U_R$) to the left (right) of the
interface boundary. Thus, roughly speaking, initially the system to the left (right) 
of the boundary surface is in the topological insulator (Mott insulator) phase. We then proceed with 
the standard DMFT iteration procedure, in which the self-energy and the cluster parameters
of the layers in the embedded region are updated according to the prescription described
in Sec. \ref{sec_2}. The iteration procedure is repeated until the local self-energy and Green's
function of each layer converge and no longer change with further iterations.

\begin{figure}[t] 
\begin{center}
\includegraphics[width=7.5cm,height=10.0cm]{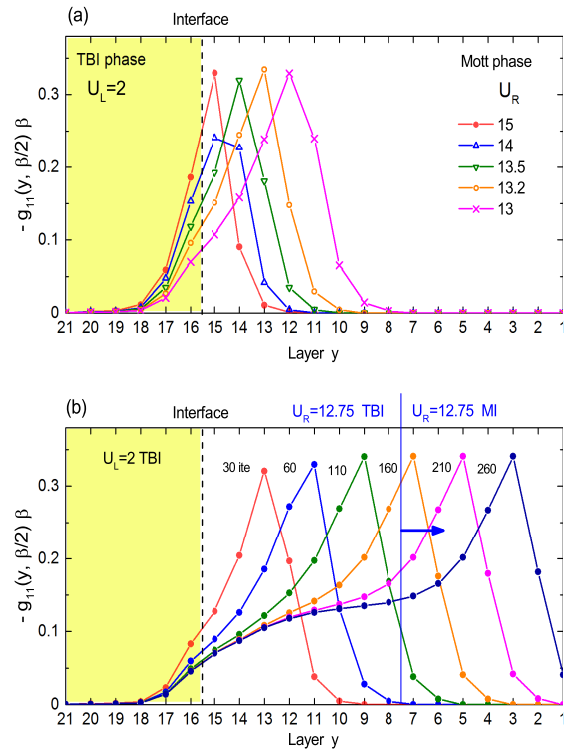}
\end{center}
\caption{\label{Fig7} (Color online)
(a) Edge state at interface between topological band insulator (left) and Mott insulator (right).
Shown is the partially integrated density of states $-\beta g_{11}(y,\beta/2)$ of orbital $\alpha=1$
as a function of layer index. The embedding region consists of 21 layers: 6 layers at $U_L=2$ and
15 layers in the range $U_R=13 \ldots 15$. Outside the embedding range, bulk behavior is assumed.
The location of the edge state is seen to be a sensitive function of the Coulomb energy in the Mott
insulator. The maximum value $-\beta g_{11}(\beta/2)\approx 0.33$ is associated with the Kondo peak.
(b)
Amplitude of the edge state as a function of layer index at $U_R=12.75$ for increasing numbers of
iterations in the self-consistency procedure. The maximum due to the Kondo peak is seen to shift 
toward the right-hand side of the embedding region.
The solid vertical line locates the phase boundary between topological and Mott insulating phases,
with both having the same Coulomb repulsion $U=12.75$, at the 160th DMFT iteration.
}\end{figure}

Figure\ \ref{Fig7}(a) shows the partially integrated DOS $-\beta g_{11}(y,\beta/2)$ at $\mu=0$
for several values of $U_R$ as a function of layer index $y$. This quantity exhibits a prominent
maximum near the interface which is associated with the edge state appearing at the
phase boundary between the topological and Mott insulators.
The edge state for $U_R=15$ is seen to be well localized at the interface. The Mott gap in the
right-hand system at this Coulomb energy is rather large, so that the edge state decays rapidly
into the Mott insulator. In the band insulating system on the left side, the gap is also large,
so that the shape of the edge state in this region is similar to the one at the insulator-vacuum
interface shown in Fig.\ \ref{Fig3}.

\begin{figure}[t] 
\begin{center}
\includegraphics[width=8.8cm]{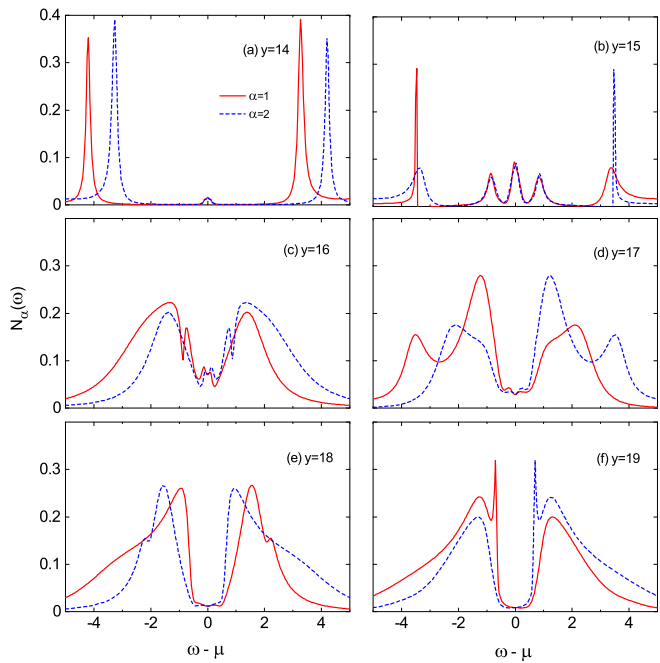}
\end{center}
\caption{\label{Fig8} (Color online)
(a) Interacting DOS,  $N_{\alpha}(\omega)$, at interface between topological band insulator
($U_L=2$) and Mott insulator ($U_R=15$). Solid (red) curves: orbital $\alpha=1$, dashed (blue) curves:
orbital $\alpha=2$. Panels (a) and (b) correspond to two surface layers on the right of the interface
($y=14,\,15$ in Fig.\ \ref{Fig7}),
panels (c) to (f) to four surface layers on the left of the interface ($y=16\ldots19$).
These spectra are obtained by extrapolating the lattice Green's function to $\omega+i\gamma$
with a small imaginary energy $\gamma=0.05$ via the routine {\it ratint}. The central spectral feature 
in panel (b) corresponds to the Kondo peak induced in the Mott insulator via the usual proximity effect 
due to the metallic edge state, while in panel (c) the low-energy feature is induced in the surface
layer of the topological insulator via a reverse proximity effect caused by the Kondo peak. 
}\end{figure}

We point out that the amplitude of $-\beta g_{11}(y,\beta/2)$ in the first layer of the Mott insulator 
for $U_R=15$ is significantly larger than in the surface layer of the topological band insulator. 
This enhancement is related to the fact that, as a result of a proximity effect, a Kondo 
resonance appears in the Mott insulator due to the screening of the localized spins via the 
helical edge states.\cite{Ueda:13} Thus, the interface may be viewed as a Kondo-lattice, where 
the metallicity is associated with the edge state induced by the topological band insulator.
This is illustrated in 
Fig.~8, which shows the variation of the interacting DOS with layer index for $U_L=2$, $U_R=15$.
In the first layer of the Mott insulator (panel (b)), the DOS at low energies has a three-peak 
structure, consisting of Kondo peak and van-Hove singularities at the limits of the edge state.
The maximum of $-\beta g_{11}(y,\beta/2)\approx 0.33$ at $y=15$ (see Fig.~7(a)) can therefore serve as a signature
of the Kondo peak. In the second layer, only a weak remnant of this peak is observed. In contrast, 
the first layer of the band insulator is dominated by the van Hove features of the one-dimensional 
metallic edge state. The deeper layers reveal the appearance of the bulk band gap, in close 
correspondence to the behavior at the bare surface shown in Fig. 5.         
   
As can be seen in Fig.\ \ref{Fig8}(c), the DOS near $\omega=\mu$ in the surface layer of the
topological insulator is enhanced due to the presence of the Mott insulator. This is also
evident by comparing $-\beta g_{11}(y,\beta/2)\approx 0.12$ for $U=2$ at the free surface
(see Fig.\ \ref{Fig3}) with the corresponding value ($\approx 0.18$) in the surface layer for 
$U_L=2$ (Fig.\ \ref{Fig7}(a)). Thus, the Kondo peak in the Mott insulator gives rise, via the 
single-particle hopping across the interface, to a low-energy spectral feature at the surface 
of the topological band insulator. Note that this feature is also present for $U_L=0$. It is 
therefore not induced by the small local self-energy in the band insulator, but by the large    
self-energy in the neighboring layer exhibiting the Kondo peak. This mechanism may therefore be 
viewed as a 'reverse proximity effect', in contrast to the usual one, in which the metal
states induce the Kondo peak in the Mott insulator. This kind of 'feedback' effect occurs also    
at interfaces between ordinary metals and Mott insulators (see, for instance, the small peak 
for $x=0$ in Fig.\ 1 of Ref.\ \onlinecite{Helmes}). In the latter case, however, this effect 
is very small because of the dominant metallic DOS. In the present heterostructure, this effect 
is much more pronounced because of the minimum of the density of states of the edge state 
in the surface layer of the topological band insulator.

So far we have discussed the formation of the Kondo peak for $U_L=2$ and $U_R=15$. Returning to   
Fig.\ \ref{Fig7}(a), we point out that, 
when the Coulomb energy $U_R$ in the right-hand system is lowered, the maximum of the edge state 
shifts away from the interface toward the interior of the Mott insulator. Evidently, due the proximity 
of the topological band insulator, the topological band insulating phase is more stable in the surface 
region of the Mott insulator, so that the effective boundary between the topological and Mott phases
moves away from the physical interface. From a numerical point of view, one observes that
the self-energy and cluster parameters of the boundary layers are converted from those
characteristic of the Mott insulator phase to those characteristic of the topological insulator
phase with increasing iterations, in a layer-by-layer fashion, starting from the first boundary layer, 
toward the interior of the Mott insulator, until no further phase conversion of layers takes place. 

Note that for the lowest two Coulomb energies shown in Fig.\ \ref{Fig7}(a)
($U_R=13.0$ and $U_R=13.2$), the Mott insulator is within the bulk coexistence region
($U_{c2}\approx 13.4$, see Fig.\ \ref{Fig2}), so that the actual phase depends sensitively on
the properties near the interface. For $U_R=13.2$, the bulk Mott insulator phase is more stable
than the bulk topological insulator phase, so that the DOS profile shown in Fig.\ \ref{Fig7}(a) does
not shift any more with further iterations. On the other hand, $U_R=13.0$ seems to be very close
to $U_b$, so that one needs hundreds of iterations to reach the DOS profile in Fig.\ \ref{Fig7}(a). 
It is to be noted that to ensure the persistence 
of the edge state at these Coulomb energies, the bulk phase in the asymptotic region on the 
right-hand side is assumed to be Mott insulating. As discussed in the previous subsection, if instead
both constituents of the heterostructure are topological band insulators, the edge state disappears. 

To illustrate this delicate balance between topological and Mott insulating solutions in the interface
region, we show in Fig.\ \ref{Fig7}(b) the edge state for $U_L=2$ and $U_R=12.75$. Beyond the 15 
surface layers of the right-hand system, bulk Mott insulating behavior is assumed. In addition, 
these 15 layers are initially assumed to be in the Mott insulating phase. With increasing number 
of  iterations, the peak of the edge state is seen to shift toward the right side of the embedded 
region. Apparently, at $U_R=12.75$, the bulk topological insulator phase is more stable than the bulk
Mott insulator phase, so that the conversion of layers from the Mott phase to topological phase does 
not stop until all layers in the embedded region are converted. In other words, for this value of 
$U_R$, one cannot find a stable Mott solution of the DMFT equation in contrast to the cases with 
larger $U_R$ shown in panel (a). The marked increase of spectral weight between layers 15 and 16 
reflects the fact the nominal boundary of the heterostructure now comprises neighboring topological
insulating phases for different values of $U$ (see Fig.\ \ref{Fig6}).        
Accordingly, the shape of the `buried' edge state within the right-hand system approaches that 
at the interface between systems with identical Coulomb energies in the coexistence range, but 
with topological band insulating and Mott insulating phases present on either side.

\begin{figure}[t] 
\begin{center}
\includegraphics[width=7.5cm,height=5.0cm]{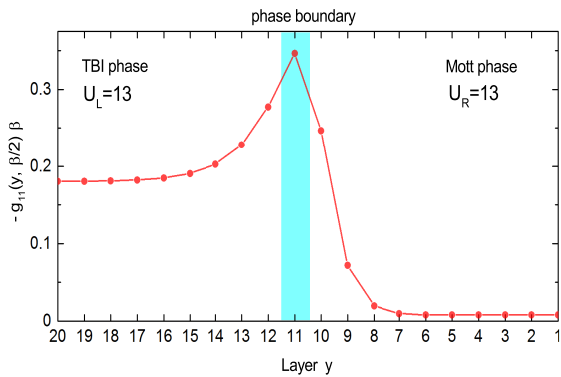}
\end{center}
\caption{\label{Fig9} (Color online)
Edge state at interface between topological band insulator and Mott insulator in the coexistence 
region with $U_L=U_R=13$. The Mott gap at this Coulomb energy is much larger than the topological
band gap. Thus, the asymptotic value of $-\beta g_{11}(y,\beta/2)$ on the right-hand side is much
lower than on the left-hand side, and the decay of the edge state within the Mott insulator is 
more rapid. The vertical bar denotes the Kondo peak in the surface layer of the Mott insulator 
at the effective boundary with the topological band insulator.       
}\end{figure}

The edge state in such a case is shown in Fig.\ \ref{Fig9} for $U_L=U_R=13$, where bulk Mott insulating 
(bulk topological band insulating) behavior is enforced on the right (left) side of the embedding 
region, respectively. The asymptotic value of $-\beta g_{11}(y, \beta/2)$ on the right-hand side is 
very small because of the large size of the Mott gap. For the same reason, the decay of the edge 
state in the Mott insulator is more rapid than within the band insulator. The spatial distribution 
of this edge state is very similar to the one in Fig.\ \ref{Fig7} deep within the nominal Mott insulator. 
As stated above, since $U_L=U_R=13$ is very close to $U_b$ at $T=0.01$, the bulk topological
insulator on the left-hand side and the bulk Mott insulator on the right-hand side have nearly the
same stability. Therefore, the DOS profile in Fig.\ \ref{Fig9} does not move with additional DMFT iterations. 

Evidently, the phenomenon observed in Fig.\ \ref{Fig7} is a proximity effect, where the band insulating 
properties on one side of the interface are induced up to a certain depth on the other side, for instance, 
via the penetration of the edge state wave function across the boundary layer, although asymptotically
this side is a Mott insulator. The boundary between band and Mott insulating phases then does not
coincide with the nominal interface.

The results discussed above suggest that the embedding scheme might be useful for the study of the 
relative stability of coexisting phases in DMFT calculations. Let us assume $U_L=U_R=U$ lies in the 
coexistence domain, with band insulating (Mott insulating) properties enforced on the left (right) 
side of the embedded region. As long as $U_{c1}(T)<U<U_b(T)$, the topological band insulating solution 
is more stable, so that the edge state will be located at the right boundary of the embedded region. 
Conversely, if $U_b(T)<U<U_{c2}(T)$, the Mott phase is more stable, so that the edge state shifts 
toward the left boundary.

\section{Summary}

The edge state at the interface between topological band and Mott insulators has been
investigated within inhomogeneous DMFT. The generalized Bernevig-Hughes-Zhang two-band
model is used to describe the interplay between interorbital hybridization and local 
Coulomb energy. The electronic properties in the vicinity of the interface are treated
self-consistently by making use of the embedding scheme, where the effect of the 
asymptotic semi-infinite bulk materials is described in terms of complex local potentials.
The finite-temperature exact diagonalization method is employed to evaluate the on-site
many-body interactions.  

The main result of this work is the observation that, close to the critical Coulomb energy of 
the correlated topological insulator, the edge state is expelled from the interface toward the 
interior of the Mott insulator. Thus, as a result of the proximity with the topological band 
insulator, the Mott insulating phase within a certain depth is converted to a topological band 
insulating phase, where the width of the conversion region depends on the local Coulomb energy 
within the Mott insulator. With increasing Coulomb energy, 
the Mott gap widens and the topological edge state is pushed again toward the interface. 
Its decay within the Mott insulator then becomes more rapid. At Coulomb energies below the 
Mott transition, the edge state ceases to exist since in this case the interface corresponds 
to that between two weakly and moderately correlated topological band insulators.

The origin of the interface-induced conversion from Mott insulating to band insulating 
behavior is the coexistence region associated with the first-order nature of the Mott 
transition. Depending on the temperature of the sample, either the Mott insulating or
the band insulating solution is more stable at a given value of the local Coulomb energy.            
These results suggest that the embedding method might be useful to determine the relative 
stability of Mott and band insulating phases in DMFT calculations. In addition, the conversion 
between Mott and band insulating phases might be observed experimentally in alloy heterostructures 
in which the existence of these phases can be tuned continuously by varying the material parameters.  

We have also shown that the normal proximity effect, where a Kondo peak in a Mott insulator is
induced via neighboring metallic states, gives rise to a reverse proximity effect, where the Kondo 
peak leads to an enhanced density of states at the surface of the neighboring metal.

\vskip3mm

\begin{acknowledgments}
The work of H. I. was supported by JSPS KAKENHI (No. 24540328). 
H.I. thanks the Alexander von Humboldt foundation for support. 
Part of the computations were carried out using the J\"ulich Juropa Supercomputer.
A. L. thanks Theo Costi for useful discussions.
\end{acknowledgments}

\end{document}